\documentclass[english,12pt]{article}
\usepackage[ruled]{algorithm2e}

\usepackage{cite}
\usepackage{amsmath, amssymb}
\usepackage{hhline}
\usepackage{alltt}
\usepackage{hyperref}
\usepackage{subfig}
\usepackage{epsfig}

\usepackage{makeidx}         % allows index generation
\usepackage{graphicx}        % standard LaTeX graphics tool
\usepackage[T2A]{fontenc}
\usepackage[utf8]{inputenc}
\usepackage{babel}
\usepackage{url}
\newcommand{\vect} \mathbf

\begin{document}

\title{A Parallel Algorithm for Calculation of Large Determinants with High Accuracy for GPUs and MPI clusters}
\author{Gleb Beliakov
\\
School of Information Technology, Deakin University, \\221 Burwood Hwy, Burwood
3125, Australia \\ \texttt{gleb@deakin.edu.au}\\ and \\
Yuri Matiyasevich
\\
St.Petersburg Department of Steklov Institute \\ of
Mathematics of Russian Academy of Sciences, \\
27 Fontanka, St.Petersburg, 191023, Russia
\\ \texttt{yumat@pdmi.ras.ru}
}
\date{}
\maketitle

%\IEEEpeerreviewmaketitle

\begin{abstract}
We present a parallel algorithm for calculating very large determinants with arbitrary precision on computer clusters.
This algorithm minimises data movements  between the nodes and computes not only the determinant but also all minors corresponding to a particular row or column at a little extra cost, and also the determinants and minors of all submatrices in the top left corner at no extra cost. We implemented the algorithm in arbitrary precision arithmetic, suitable for very ill conditioned matrices, and empirically estimated the loss of precision. The algorithm was applied to studies of Riemann's zeta function.
\end{abstract}

\textbf{Keywords} {\small determinant, linear algebra, parallel algorithms, Message Passing Interface, GPU, Riemann's zeta function.}

\baselineskip=\normalbaselineskip

%\thispagestyle{empty}

%as many elements of $\vect x$ smaller than or equal to  $x_i$ as those bigger than or equal to $x_i$, provided $n$ is odd.

\section{Introduction}

Parallel linear algebra algorithms have been developed for many hardware architectures with the aim of accelerating routine calculations \cite{Agullo2009,Haidar:2011:ADS:2306414.2306421,slug,Brent1991}. In this paper we report on our experiences with parallel calculation of determinants of large nearly singular matrices with very high accuracy.

The need to solve such a problem came from research on the famous Riemann's \emph{zeta function}.
the distribution of  zeroes
of zeta function has puzzled mathematicians for over a century. The famous Riemann Hypothesis \cite{Borwein2008_book}, which was included by D. Hilbert at the
very end of  XIX century
as part of his 8-th problem, and which is also one of the Clay Institute seven Millennium Problems \cite{millenium}, is to prove or disprove  that all non-real zeroes of $\zeta(z)$
lie on the critical line $\Re(z)=\frac{1}{2}$.
 Among other things, Riemann's zeta function is closely related to the distribution of prime numbers, whose fundamental role in mathematics is well known.

 Recently,  the second author
has proposed what he called \emph{Artless method} for studying zeroes of the zeta function (see \cite{YumatLeicester,yumatFuncan1,yumatFuncan2,yumatBonn} and visit \cite{yumatartless}).
 It consists in approximating zeroes by using a special interpolant built on already known zeroes of zeta function. In the process of  building the interpolant, it is necessary to compute with high accuracy the determinant of a large matrix, as a function of a parameter in its last column, and hence the signed minors corresponding to the last column are needed. The matrix is nearly singular, and it was required to perform calculations with the accuracy of ten thousand decimal places. Furthermore, certain patterns in the behaviour of  these minors as a function of matrix size $N$ needed to be investigated \cite{yumatartless,YumatLeicester,yumatBonn},
  and therefore the sequence of arrays of minors for matrix sizes $N=2,3,\ldots$ was required. All those special requirements prevented us from mapping directly the problem at hand to one of the standard parallel linear algebra solutions and using existing tools such as ScaLAPACK or PLASMA \cite{Agullo2009,Haidar:2011:ADS:2306414.2306421,slug}.

In this work we describe our approaches to solving the problem of calculating the required sequence of minors with high precision on various architectures. Initially we attempted using Graphics Processing Units (GPUs) for this task, and designed and implemented a parallel algorithm for this architecture, using quad-precision (256-bits) and arbitrary precision libraries \cite{GPUPREC, GPUPREC_conf}. Our initial experiments, though, indicated that GPU architecture delivered about the same efficiency as one single core of a modern CPU when using arbitrary precision libraries (with over 1000 bits accuracy). Subsequently we implemented a parallel algorithm for shared memory multicore architecture using \texttt{pthread} library, which exhibited linear reduction of CPU time when using up to 16 cores. However, it soon became clear that only a sufficiently large cluster would be capable of solving the problem with ten thousand decimal places (32Kbits accuracy) for $N$ in the order of 10 thousand. Indeed, just to store the elements of one matrix of that size, 400 Gb of RAM was needed. Therefore we designed and implemented an MPI (Message-Passing Interface) algorithm and verified it on several computer clusters, running on up to 200 cores. We dealt appropriately with the issue of load balancing, and at the end observed nearly linear gain in performance with the number of nodes used. We successfully verified the algorithm on the computations with two types of matrices needed for the Artless method.

This paper is structured as follows. In Section \ref{sec1} we will formally describe the problem and give a brief introduction to some elements of the Artless method. In Section \ref{sec2} we present computational algorithm and its various parallelizations for GPU-CUDA, pthread, OpenMP and MPI platforms. In Section \ref{sec3} we detail some of the numerical experiments and quantify the performance of our algorithm, including the loss of accuracy. Section \ref{sec_conc} concludes.

\section{Motivation and problem formulation} \label{sec1}

%\subsection{Motivation}
  Zeta function is defined in the complex plane by the analytic continuation of the series
\begin{equation}\label{zeta}
    \zeta(s)=\sum_{n=1}^\infty n^{-s},
\end{equation}
which converges for $\Re(s)>1$. It satisfies the functional equation
\begin{equation}
\zeta(s)=2^s \pi^{s-1}\sin\left(\frac{\pi s}{2}\right)\Gamma(1-s)\zeta(1-s).
\end{equation}
Riemann also defined a symmetric version of the above, using function $\xi$,
\begin{equation}
\xi(s)=\frac{1}{2}\pi^{-s/2}s(s-1)\Gamma\left(\frac{s}{2}\right)\zeta(s),
\end{equation}
which yields a simpler functional equation
\begin{equation}
\xi(s)=\xi(1-s).
\end{equation}

The \emph{trivial zeroes} of $\zeta$ are negative even integers, and they are its only real zeroes.
The other, \emph{non-trivial zeroes} can only be found in the strip $0<\Re(s)<1$. The zeroes of the function $\xi$ are exactly the non-trivial zeroes of the function $\zeta$. Also following Riemann, we make a change of variables $s=\frac{1}{2}+it$ and define
\begin{equation}\label{Xi}
    \Xi(t)=\xi\left(\frac{1}{2}+it\right).
\end{equation}
The functional equation implies that $\Xi$ is even function, $\Xi(t)=\Xi(-t)$.
With this notation, the Riemann Hypothesis states that all zeroes of $\Xi$ are real numbers.

An important relation of zeta function to prime numbers was given by von Mangoldt. Let Chebyshev function $\psi$ be
$$
\psi(x)=\sum_{q \mbox{ is a power of prime }p}^{q \leq x} \ln(p)=\ln(\mathrm{LCM}(1,2,\ldots,\lfloor x\rfloor)).
$$
Then for non-integer $x$ greater than~$1$, by von Mangoldt's Theorem $\psi(x) $ can be expressed as
\begin{equation}
\psi(x)=x-  \sum_{n=1}^\infty \frac{x^{-2n}}{-2n}-\sum_{\xi(\rho)=0}\frac{x^\rho}{\rho}-\ln(2n).
\label{mang}\end{equation}
The first sum runs over the trivial zeroes of zeta,
 and the second sum runs over the non-trivial zeroes.
In fact, knowing zeroes of zeta function, one could compute primes by merely looking at  the graph of the right-hand side of \eqref{mang},
 and identifying the powers of primes by jumps in the graph.

Assuming the Riemann Hypothesis and  that all zeroes are simple, let us denote by $\pm \gamma_1, \pm \gamma_2,\ldots$ the real  zeroes of $\Xi$ listed by increasing of absolute values.

Further, let
\begin{equation}
\beta_n(t)=-\frac{\pi^{-\frac{1}{4}+\frac{it}{2}}(t^2+\frac{1}{4})
\Gamma(\frac{1}{4}-\frac{it}{2})}{4n^{\frac{1}{2}-it}}-
\frac{\pi^{-\frac{1}{4}-\frac{it}{2}}(t^2+\frac{1}{4})
\Gamma(\frac{1}{4}+\frac{it}{2})}{4n^{\frac{1}{2}+it}}.
\end{equation}
Thanks to the functional equation we formally have
\begin{equation}\label{seriesbeta}
\Xi(t)=\sum_{n=1}^\infty \beta_n(t).
\end{equation}

 We define the \emph{interpolating determinant} as
\begin{equation}
    \Delta_N(t)=\left|
                         \begin{array}{cccc}
                           \beta_1(\gamma_1) & \ldots & \beta_1(\gamma_{N-1})  & \beta_1(t)  \\
                           \vdots & \ddots & \vdots & \vdots \\
                           \beta_N(\gamma_1) & \ldots & \beta_N(\gamma_{N-1}) & \beta_N(t) \\
                         \end{array}
                            \right|.
\label{det}\end{equation}

Clearly, the determinant $\Delta_N(t)$ vanishes as soon at $t$ is equal to
$\pm \gamma_1,\dots, \pm \gamma_{N-1}$, because for such a  $t$ there are
 two equal columns in \eqref{det}.
The surprising observation is that $\Delta_N(t)$ vanishes also at certain points extremely close to
a number of the next zeroes $\pm \gamma_{N},\dots, \pm \gamma_{N+k}$.

  The larger is $N$, the more
   subsequent zeroes approximately coincide with the zeroes of
   $\Delta(t)$. For example determinant $\Delta_{3000}(t)$ has
   zeroes having more than 500 common decimal places with $\gamma_{3001},
   \ldots, \gamma_{3020}$.

The determinant $\Delta_N(t)$ can be expanded into the linear combination of
functions $\beta_n(t)$,
\begin{equation}
\Delta_N(t)=\sum_{n=1}^N \tilde \delta_{N,n} \beta_n(t),
\end{equation}
where
\begin{equation}\label{minorsdef}
\tilde \delta_{N,n}=(-1)^{N+n} \left|
                         \begin{array}{ccc}
                           \beta_1(\gamma_1) & \ldots & \beta_1(\gamma_{N-1})   \\
                           \vdots & \ddots & \vdots \\
                            \beta_{n-1}(\gamma_1) & \ldots & \beta_{n-1}(\gamma_{N-1}) \\
                            \beta_{n+1}(\gamma_1) & \ldots & \beta_{n+1}(\gamma_{N-1}) \\
                             \vdots & \ddots & \vdots  \\
                           \beta_N(\gamma_1) & \ldots & \beta_N(\gamma_{N-1}) \\
                         \end{array}
                            \right|.
\end{equation}

The object of interest are normalized minors
\begin{equation}\label{minors}
    \delta_{N,n}=\frac{\tilde \delta_{N,n}}{\tilde \delta_{N,1}}.
\end{equation}
In particular, function
$$
\sum_{n=1}^N \delta_{N,n} \beta_n(t)
$$
has the same zeroes as $\Delta_N(t)$.

The second matrix of  interest  had a slightly different structure
\begin{equation}\label{betaNew}
    \tilde \Delta_N(t)=\left|
                         \begin{array}{cccccc}
                          1 & 1& \ldots & 1 & 1 & 1  \\
                          \vdots & \vdots & \ddots &\vdots & \vdots &\vdots\\
                           n^{-\bar \rho_1} & n^{-\rho_1} & \ldots & n^{-\bar \rho_M} & n^{-\rho_M}  & n^{-\frac{1}{2}-it} \\
                          \vdots & \vdots & \ddots &\vdots & \vdots &\vdots\\
                           N^{-\bar \rho_1} & N^{-\rho_1} &\ldots & N^{-\bar \rho_M} & N^{-\rho_M}  & N^{-\frac{1}{2}-it}
                         \end{array}
                            \right|,
\end{equation}
where $\rho_n=\frac{1}{2}+i \gamma_n$ are the non-trial zeroes of $\zeta$ and $N=2M+1$. Similarly, zeroes of $\tilde \Delta_N(t)$ for $N=3001$ have more than 1000 common decimal places with $\gamma_{1501}, \ldots, \gamma_{1561}$.

The second author observed
several patterns of behaviour of the sequences of coefficients
$\delta_{N,n}, n=1,\ldots,N$ for various $N$ \cite{yumatartless}. These patterns have clear number-theoretical
meaning, but in order to
discover and assess such patterns, which have very fine
structure, one has to compute the minors given in (\ref{minors})
with extremely high accuracy  to avoid numerical artifacts.
Further, the interpolating matrices are nearly singular, and  the
values $\delta_{N,n}$ approach zero very rapidly with $N$
(e.g. $\delta_{100,1} \approx 10^{-120}$),
 so one expects very large losses of precision in numerical calculation. All this dictates the need to use very high accuracy in  calculations, of order of ten thousand decimal places.

At this point we abandon the topic of Riemann's zeta function, and focus squarely on calculating the determinants $\Delta_{N}$ (or $\tilde \Delta_{N}$) and the corresponding normalized minors, as the main numerical complexity of the Artless method comes from these calculations.

\section{Detailed algorithm} \label{sec2}

There is a number of ways determinants can be evaluated numerically. Matrix factorisation is a standard method, although not necessarily the best when computations are parallelised. Our initial thought was to look at parallel versions of the condensation method. This condensation method was inspired by the celebrated Dodgson's\footnote{C.T. Dodgson is better known as Lewis Carroll.} condensation method \cite{Dodgson}. In Dodgson's method, an $N \times N$ matrix determinant is ``condensed" into an $N-1 \times N-1$ determinant by calculating $N^2$ connected subdeterminants of sizes $2 \times 2$. This algorithm is trivially parallel, yet it suffers from the requirement of having no zeroes in the interior of the  matrix.

Inspired by the parallel nature of the condensation steps, the authors of \cite{condensation_arxiv} proposed a variant of the condensation method, in which the determinant of an $N \times N$ matrix $A$ is replaced by the determinant of an  $N-1 \times N-1$ matrix $B$, calculated as follows
\begin{equation}\label{condensation}
b_{ij}=\left|\begin{array}{cc}
               a_{1,l} & a_{1,j+l} \\
               a_{i+1,l} & a_{i+1,j+1}
             \end{array} \right|
\end{equation}
for $j\geq l$, and $b_{ij}=-a_{i+1,j}a_{1,l}$ for $j<l$, where $l$ is the smallest column index of a non-zero element in the first row of $A$. The relation between $A$ and $B$ is that
\begin{equation}
\det(A)=\det(B)/(a_{1,l})^{N-2}.
\end{equation}
After $N-1$ condensation steps the determinant of $A$ reduces to a single number divided by a product of powers of pivot factors.

The condensation method from \cite{condensation_arxiv} was taken up by \cite{Moreno_2012}, where it was slightly modified to avoid division by $(a_{1,l})^{N-2}$, by factoring out the pivot element $a_{1,l}$ from the first column of $A$ (hence getting a modified matrix $\tilde A$), so that the relation between the determinants of the matrices becomes
\begin{equation}
\det( A) = a_{1,1} \det(\tilde A)=a_{1,1} \det(\tilde B),
\end{equation}
with $\tilde B$ obtained from  $\tilde A$ in  (\ref{condensation}).

Successive condensation steps produce the array of pivot elements, whose product is the determinant of $A$. This modification of the condensation method is data parallel as elements of $B$ are computed independently of each other.
The authors of \cite{Moreno_2012} developed a parallel version of their algorithm suitable for GPUs and assessed its performance.

We started with formula (\ref{condensation}) and also developed our own GPU parallel algorithm, differing from that of \cite{Moreno_2012} in that all computations were performed in place (instead of having two rotating matrices for input and output), and that we used high precision arithmetic for all steps (we used quad-precision QD library for GPUs and GPUPrec arbitrary precision libraries for GPUs \cite{GPUPREC,GPUPREC_conf}). We soon realised though, that condensation steps in (\ref{condensation}) were exactly the steps of Gaussian elimination, and hence the version of condensation algorithm in \cite{condensation_arxiv} and \cite{Moreno_2012} was in fact not different from Gaussian elimination without pivoting. Optional pivoting (as it was done in \cite{Moreno_2012}) can be trivially added.

Knowing the equivalence of \eqref{condensation}  with the Gaussian elimination steps, calculations could be streamlined, and computational complexity can be confirmed to be $O(N^3)$. However, the question of  calculation of minors corresponding to the last column of the matrix $A$ was still outstanding.
The naive approach by calculating $N$ determinants (\ref{minorsdef}) was not appealing because of computational cost.

It is known that one can compute a determinant of size $N$ and all its $(N-1)\times(N-1)$ minors in the same time $O(N^3)$. It is based on the result from \cite{Strassen} which establishes that the complexity of evaluation of all partial derivatives of certain functions (in particular multivariate polynomials) is a constant multiple of the complexity of evaluating the function itself. Since all the $(N-1)\times(N-1)$ minors are partial derivatives of the determinant, the result follows.
%, although we were unable to find an explicit algorithm for this task.

Additionally, would it be possible to compute not just one array of signed minors $\delta_{N,n}$ for a particular $N$, but the whole series of such arrays for $N=2,3,\ldots$ in one non-redundant computation?

To answer these questions we used the following trick. Let us construct the augmented matrix
$$
C=[A \quad I],
$$
where  $I$ is the identity matrix of  size $N$. Perform Gauss elimination on $C$. For completeness, the algorithm is shown on Fig. \ref{alg:gauss}. The last $N$ elements of the last row will contain the signed minors of $A$ corresponding to the last column, divided by $|A|$. To see this, recall that Gauss elimination (without pivoting) of a non-singular matrix $A$ is equivalent to multiplying it by a lower triangular matrix $L$ from the left,
$
L A=U,
$
where $U$ is in row-echelon form. The same operations applied to $I$ result in $L I=L$. Now, the product of the last row of $L$ and the last column of $A$ yield $L_{N,\cdot} \; A_{\cdot,N}=u_{NN}=1$, since $U$ is in row-echelon form.
If we now multiply $L_{N\cdot}$ by $|A|$, we obtain the result, $(|A| L_{N,\cdot}) \; A_{\cdot,N}=|A|$.

We can see that all the required signed minors can be computed automatically using Gaussian elimination (the determinant of $A$ is found from the product of diagonal elements of $L$) with the help of additional row operations. The computational cost of it is twice that of Gaussian elimination, and computational complexity remains $O(N^3)$.

Storagewise, the matrix $L$ can be stored in the lower triangular part of $A$, because once the  row $n$ is processed in Gaussian elimination, the elements $a_{ij}$ below diagonal with $j<n$ are no longer required (indeed they are zero in row-echelon form). On the other hand, at step $n$ of the elimination process, only the elements $l_{ij}$ with $j<n$  and the main diagonal are different from zero. Therefore, nonzero elements $l_{ij}$ can be stored below the diagonal of $A$, and the only extra storage required is for the overlapping diagonals of  $A$ and $L$. Therefore  all calculations can be done in memory $1\cdot N^2+O(N)$,
as only one extra array of size $N$ is required. %Hence the question of calculating the minors with the same order of complexity as that of a single determinant is answered affirmatively.

Next we turn to the second question of whether it is possible to compute the whole series of the minors, for different sizes of the matrices $N=2,3,\ldots$. The answer here is also affirmative, and in fact no extra work or storage is required at all. To see this, note that at step $n$ of Gaussian elimination, the row $n$ of matrix $L$ contains the desired (up to a factor) minors, corresponding to the $n$th column of $A$. We apply the same reasoning as before to show this is the case: note that after step $n$, the diagonal element of the row echelon form $u_{nn}=1$, which is the product of the $n$th row of $L$ and $n$th column of $A$.

To put it into the context of our motivating problem, all
normalised minors $\delta_{N,n}$ (\ref{minors}),
$n=1,2,\ldots,N$, $N=2,3,\ldots,\bar N$ can be computed
with CPU cost $O(\bar N^3)$ and storage $O(\bar N^2)$ using
Gaussian elimination.

There are methods for matrix inversion and calculation of
determinant based on fast matrix multiplication, e.g.
Sch\"onhage-Strassen \cite{Strassen1971} and Coppersmith-Winograd
algorithms \cite{Coppersmith1990}, which have complexity
$O(N^{2.8})$ and $O(N^{2.376})$ respectively. Through the result of \cite{Strassen}, all   $(N-1)\times(N-1)$ minors can also be computed in the same time, although we were unable to find implementations of such algorithms in the literature.

However, it appears that the fast matrix multiplicaiton algorithms are not structured to facilitate computation of the whole sequence of arrays of minors for $N=2,3,\ldots,\bar N$ in one run. Therefore we settle on Gaussian elimination as the most efficient way of computing all the normalised minors $\delta_{N,n}$.

The question of available RAM becomes important for computations with larger $N$ of order of 10000.  As we mentioned previously, to store a matrix of that size, with 10 thousand decimal places accuracy, 400 GB of RAM is required.  One approach we pursued is to partition the matrix into blocks and use memory paging to fit the blocks into available RAM.

Let us partition the matricies into $B^2$ blocks of size $N_b \times N_b$, with $N=B \cdot N_b$. Each block is stored on a hard disk and is loaded to RAM when needed. As soon as certain blocks are in RAM, we perform all possible steps of Gaussian elimination with data in these blocks.
At each point in time at most four blocks are needed in RAM, hence RAM required is $4 N_b^2+O(N_b)$.
However, now  we do not use shared space to keep both matrix $A$ and $L$, so memorywise the algorithm uses $\frac{3}{2}N^2+O(N)$ space.

The operations within blocks were executed in parallel in different threads on a single host, using OpenMP library, however distinct sets of blocks could be processed in parallel on different hosts in a specified order but otherwise asynchronously. This way a cluster of hosts can be used for parallel execution.

% Algorithm
\begin{algorithm}[t]
\SetAlgoNoLine
\KwIn{$N$, $A \in R^{N \times N}$}
\KwOut{$Det$, $L$ containing signed minors}
\begin{enumerate}
    \item[1] $Det=1$, $L=I_{N}$ (identity matrix)
  \item[2] For $i=1,\ldots,N$ do:
  \begin{enumerate}
    \item[2.1] $Det=Det \cdot a_{ii}$
    \item[2.2] For $j=i+1,\ldots,N$ do:
      \begin{enumerate}
        \item[2.2.1] $z =a_{ji}/a_{ii}$
        \item[2.2.2] For $k=i+1,\ldots,N$ do: \\
        $a_{jk}=a_{jk}- z \cdot  a_{ik}$
        \item[2.2.3] $z =l_{ji}/a_{ii}$
        \item[2.2.4] For $k=1,\ldots,i-1$ do: \\
        $l_{jk}=Det \cdot (l_{jk}- z \cdot  l_{ik})$
      \end{enumerate}
   \end{enumerate}
  \item[3] return $Det$ and $L$.
\end{enumerate}
\caption{Calculation of determinant  and signed minors by Gaussian elimination algorithm. }
\label{alg:gauss}
\end{algorithm}

\section{Numerical evaluation} \label{sec3}

\subsection{Implementations on different platforms and parallelisation}

We implemented a variant of Gaussian elimination with high accuracy on four platforms:

\begin{enumerate}
  \item On GPU using CUDA and CUMP \cite{CUMP} arbitrary precision and GQD \cite{GPUPREC} quad-double (256 bits) precision libraries, connected to a single host;
  \item On a cluster of multicore CPUs using \texttt{OpenMP} for multithreading and GMP \cite{GMPLIB} arbitrary precision library, partitioning the matrix into blocks and using memory paging;
\item On a multicore CPU using \texttt{pthread} for multithreading and GMP \cite{GMPLIB} arbitrary precision library;
  \item On a computer cluster using MPI, and using GMP, MPFR \cite{GMPLIB,Fousse:2007:MMP} and MPIGMP \cite{MPIGMP} libraries.
\end{enumerate}

The following parallelisation strategies were adopted.
On GPU, at every step $i$ of the Gauss elimination algorithm, we spawned $N$ threads, and each thread $j$ was performing calculations in the $j$th column of matrix $A$ starting from row $i$. While this was efficient in terms of  coalescent global memory access pattern for data type \texttt{double} and to some degree for \texttt{quad-double}, for arbitrary precision numbers the way GPU threads accessed elements of the matrix $A$ did not matter, as the data were stored in non-sequential locations anyway, and hence was misalligned. We did not observe any differences in CPU time when parallelisation was done columnwise, rowwise or elementwise.

When we used memory paging and matrix partitioning in the second approach, parallelization was performed in two ways. First, we created a queue of tasks, from which  tasks were  taken by  idle hosts based on certain pre-conditions. Each task consisted of a block in which Gaussian elimination was performed, and pre-conditions ensured that a block was processed only when all the  blocks containing rows and columns with smaller indices have been processed. Second, within each block the computations were parallelized with OpenMP.

Common to the remaining  implementations, we parallelised the algorithm rowwise, that is, each thread was performing calculations in a specific row, or a group of rows. The simplest way is to break the matrix into $T$ blocks of rows sequentially, where $T$ is the number of threads, and let each thread process its own block. Some load balancing strategy is needed, as the first thread will complete processing of its rows before the rest and will be then idle, then it will be the second thread and so on. To avoid this, on shared memory architectures we reassigned the blocks to each thread every now and then, so that at all steps the sizes of the blocks were approximately the same. This was easily done by simply changing the pointers to blocks of rows each thread was responsible for, hence no copying of data took place, and overheads were negligible.

On distributed memory architectures, though, such as clusters running MPI, this method was not suitable as it would involve copying large chunks of data between the hosts. Instead, we interleaved the rows of the matrix, i.e., thread $j$ was processing rows $j$, $j+T$, $j+2T$,... and so on, $j=1,\ldots,T$. The threads processed the pivot rows of the matrix $A$ in turns, in round robin fashion. This way load balancing was implicit, and all threads had equal job to do until the very end of the computations. Each pivot row was broadcast to all the threads at every step $i$ of the algorithm, hence $N$ rows were broadcast altogether, which is the smallest number when the matrix is partitioned among the threads, and complexity of the data transfers was therefore $O(N^2 \log T)$ (we remind that broadcast in MPI has logarithmic time).  The row being broadcast was packed into a binary array using \texttt{MPIGMP} library \cite{MPIGMP}, and then unpacked at the receiving end.

Here we make an observation regarding the value of partitioning the matrix into blocks and using paged memory as opposed to storing the whole matrix in (combined) RAM.  It has two main advantages: 1) less RAM is needed, and 2) the upper limit on the matrix size $\bar N$ needs not be fixed in advance: we can gradually get determinants of sizes $N_b, 2 N_b, 3 N_b, \ldots$. Extra blocks can be added later on. On the other hand, less RAM means that some blocks were processed sequentially rather than in parallel. The cost of data transfer between hard disk and RAM was not significant compared to the cost of the actual computations. However the overall storage requirement was 50\% higher, as the matrix $L$ could not share the same space as $A$.

We opted for the fourth method in our final computations, based on MPI and using combined RAM of the available hosts, for the following reasons: 1) CPU and RAM resources available to us were sufficient to perform computations for $\bar N=12000$ with 10 thousand decimal places accuracy, and in fact the amount of combined RAM exceeded our permanent storage quota; 2) coding the algorithm was more straightforward; 3) at that moment in time, we did not need to increase $\bar N$ beyond 12000, because our computations have shown that the loss of accuracy during numerical computations was such that the results were unreliable beyond that number.

\subsection{Evaluation of efficiency}

We compared the speed of determinant calculations on various platforms using the following hardware:
Tesla C2070 GPUs with 6Gb of RAM, and clusters of Intel  E5-2670 nodes with $48-64$ Gb of RAM, connected by 4x QDR Infiniband Interconnect, running CentOS 6 linux.
The hardware was provided by the VPAC and Monash e-research centre \url{http://www.vpac.org}, \url{http://www.monash.edu.au/eresearch/}.

While Tesla GPU was quite efficient in evaluating determinants with GQD package, when we passed to much higher accuracy (from 256 to 8192 bits and more), GPU performance became equivalent to that of one CPU core, see Table \ref{tab1}. The computing time grew with the accuracy at the rate of $m^2$, $m$ being the number of bits used for multiprecision float numbers. This is consistent with the schoolbook multiplication algorithm employed in CUMP library.
%It seems little advantage could be taken from multiple GPU cores, as multiprecision arithmetic ...

Subsequently we implemented a parallel shared memory algorithm using \texttt{pthread} library, and ran it on up to 16 Intel CPU cores (on the same host). The efficiency increased linearly with the number of cores employed, as no significant interprocess communication took place, see Table \ref{tab2}. Therefore the CPU time has decreased 16-fold compared to running the algorithm on a GPU. Furthermore, since GMP library \cite{GMPLIB} uses a faster Karatsuba multiplication algorithm with complexity $O(m^{\log_2 3})\approx O(m^{1.585})$, the benefit was even greater.

As we mentioned previously, current shared memory architecture has a limitation of the total RAM that could be used and the limited number of cores. Therefore we opted to parallelising the algorithm for an MPI cluster. Here we observed linear gain in performance as evidenced by Table \ref{tab3}. This is consistent with the fact that the dominating term in the overall complexity of the algorithm  $O(N^3)$ is due to computations, and the complexity of data transfers  $O(N^2 \log T)$ is comparatively small.

\subsection{Loss of accuracy} \label{secacc}

Given that the matrices involved in the computations are large and  ill conditioned, a valid question of the accuracy of the end result arises. To estimate the loss of accuracy we performed computations with various precisions: 8 Kbits, 32 Kbits, and to a limited extent, 64 Kbits. The input data, the zeroes of Riemann's zeta function were taken with just over 32 Kbits accuracy (10000 decimal places) \cite{AccurateZeros}. By comparing the results calculated with different accuracies, we could estimate empirically the rate of accuracy loss during Gaussian elimination.

\begin{figure*}[htpb]
\begin{center}
\includegraphics[height=3in]{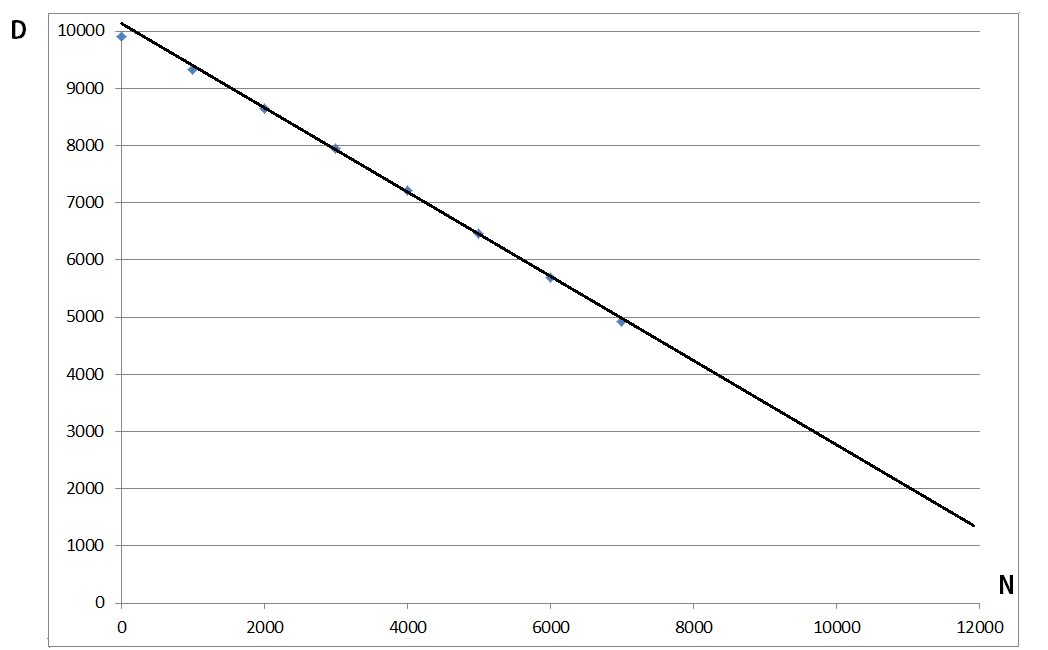}
\end{center}
\caption{Recorded loss of accuracy (decimal places vs matrix size $N$) and its linear approximation (solid line), for calculations with 32 Kbits accuracy and determinant $\tilde \Delta_{N}$ in (\ref{betaNew}).} \label{accuracy64}
\end{figure*}

\begin{figure*}[htpb]
\begin{center}
\includegraphics[height=3in]{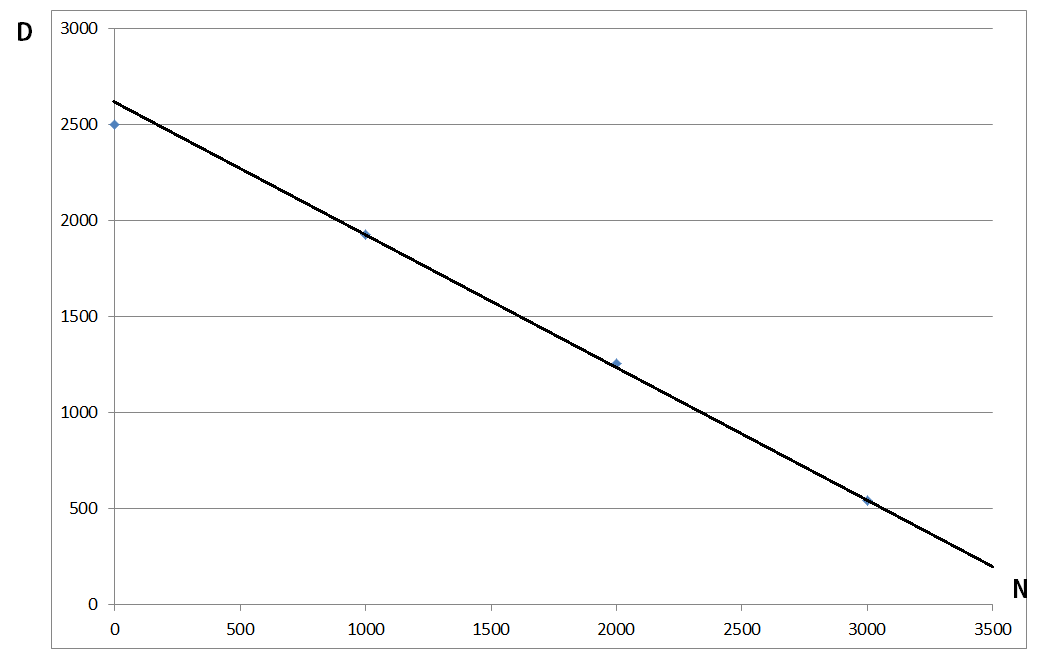}
\end{center}
\caption{Recorded loss of accuracy (decimal places vs matrix size $N$) and its linear approximation (solid line), for calculations with 8 Kbits accuracy and determinant $\tilde \Delta_{N}$ in (\ref{det}).} \label{accuracy8}
\end{figure*}

Figures \ref{accuracy64}, \ref{accuracy8} present  graph of the number of coinciding decimal places in the normalised minors $\tilde \delta_{N,n}$ as a function of matrix size $N$. We compared both the average and the smallest number of coinciding decimal places for $1\leq n \leq N$, which turn out to be very close (difference $ <2$\%).

What we observe is that the accuracy decays linearly. The equation of regression line in Figure \ref{accuracy64} is
$D=-0.72 N + 10020$.  This indicates  that the accuracy of calculations with 32 Kbits accuracy is predicted to have 1400 correct decimal places. On the other hand, calculations with 8 Kbits accuracy were valid only up to $N=7000$, and after that point were completely unreliable. Therefore the choice of 10000 decimal places accuracy was sufficient, and also necessary for calculations with matrices up to $N=12000$ in size. The coefficient $-0.72$ indicates the number of decimal places lost in each row operation. It corresponds to $0.72 \log_2 10 \approx 2.4$  bits average accuracy loss per one row operation.
This result also shows that if the input values were calculated with 10000 decimal places, then it only makes sense to increase $N$ to 13500, because after that size the loss of accuracy will make the results unreliable.

The above analysis was performed for determinants $\tilde \Delta_{N}$ in (\ref{betaNew}). A similar analysis was performed for $ \Delta_{N}$ in (\ref{det}), which yielded a slightly different regression lines, e.g.\ the loss of accuracy at $8 $ Kbits was approximated by $D=-0.42 N + 2690$. The coefficient $-0.42$ indicates that for that matrix, the loss of accuracy was $0.42 \log_2 10\approx 1.4$ bits for each row operation. Hence it appears that matrix in (\ref{det}) is slightly better conditioned than that in (\ref{betaNew}).

\subsection{Short-term outcomes}

After completing initial evaluations, we performed two production runs of the algorithm on VPAC (\url{http://www.vpac.org}) and MASSIVE (\url{http://www.massive.org.au}) clusters, using up to 168 processes. We used the upper limit on the size of the matrix of $\bar N=12000$ and accuracy of 32768 bits. One cluster was used for $\Delta_N$ from (\ref{det}) and the second for $\tilde \Delta_N$ from (\ref{betaNew}). The accurate values of Riemann's zeroes were precomputed  \cite{AccurateZeros}.

 The algorithms ran for 5 and 7 days respectively.

The results of our computations constituted almost 700 GB of
compressed high accuracy data. Our results helped to confirm
earlier calculations in \cite{YumatLeicester} for $\Delta_N$ up to
$N=3875$, which were performed with block partitioning of the matrix $A$. However we noticed some deviations from the earlier
results for larger $N$. It was later confirmed that our current
results were correct, hence these calculations were valuable in
correcting previous computational errors. The results with the
determinant $\tilde \Delta_N$ from (\ref{betaNew})
were new, and they
helped establish various patterns in the values of the normalised
minors $\delta_{N,n}$, in particular their fine structure related
to prime numbers, as presented recently in \cite{yumatBonn}.

\subsection{Future work}

As we discussed earlier, the loss of accuracy in arithmetical operations was quite significant due to nearly degenerate matrices. In addition, matrix entries themselves were computed up to 10 thousand decimal places; that was the accuracy of zeroes of zeta function we started with. However, our analysis is empirical, it provides only an indication, but not a guarantee, that the results are accurate to the indicated number of decimal places.

Interval arithmetic can be employed to obtain rigorous results. At every operation,  the upper and lower bounds on the result can be computed. A library \texttt{MPFI} \cite{MPFI} is available for such calculations with arbitrary precision. However the cost of such computations is doubling the CPU time and memory requirements. In addition, interval computations provide pessimistic error bounds, much larger than the actual errors. A posteriori interval analysis \cite{Yumaterror} provides more realistic error bounds, but it requires $O(N^3)$ memory, and hence bears prohibitive cost in our case. It might be possible to reduce its cost by not keeping all intermediate values, and in addition it may help save CPU time, as due to the inevitable loss of accuracy, later iterations of the algorithm need not be performed with full precision. We leave this analysis for future work.

\section{Conclusion} \label{sec_conc}

Calculation of matrix determinant is one of the standard linear algebra operations needed for many computational tasks. Determinants of ill-conditioned matrices are a particular challenge because of rapidly degrading accuracy, hence very high precision calculations are needed.
We have presented an algorithm based on Gaussian elimination which computes not only the determinant but a series of determinants and minors corresponding to one column or row for matrix sizes $N=2,3,\ldots,\bar N$ in one run. The algorithm has $O(\bar N^3)$ CPU and $O(\bar N ^2)$ memory complexity.

We parallelized and implemented this algorithm on four different parallel architectures, including GPU and MPI-based clusters, using arbitrary precision arithmetics. We confirmed that the  CPU wall time decreases linearly with the number of CPU cores used (for fixed matrix size and accuracy).
Our production runs involved up to 168 cores, 400 GB combined RAM and determinants of up to $\bar N=12000$ in size with 10 thousand decimal places accuracy. We empirically estimated the losses of accuracy during Gaussian elimination and found that the accuracy used was appropriate for computations with that matrix size.

We applied the algorithm to studies of the Riemann
zeta function as part of the 8-th Hilbert Problem. The results helped the second author to observe various theoretically interesting patterns in the values of the normalized minors related to prime numbers \cite{yumatBonn}.

\section*{Acknowledgements} The authors wish
 to acknowledge support by Victorian Partnership in Advanced Computing (VPAC) and Monash e-research centre for providing computing resources at their clusters, and specifically Mr. S. Michnowicz for his help in developing MPI parallelization code. Part of the calculations was performed on the “Chebyshev” supercomputer of Moscow State
University Supercomputing Center. The work of the second author was partly supported
by the programme of fundamental research ``Modern problems of theoretical mathematics” of the  Mathematics Branch of the Russian Academy of Sciences.

\bibliographystyle{plain}
%\bibliography{median1}

\renewcommand{\baselinestretch}{1}
\begin{table}[!p]
\begin{center}
\caption{The mean execution time (sec) of Gauss elimination algorithm on a Tesla C2070 GPU using arbitrary precision library \texttt{cump} \cite{CUMP}. One can observe that the CPU time increases by about 3.95 when doubling the accuracy, consistent with the schoolbook multiplication algorithm. } \label{tab1}
\begin{tabular}{||r||r|r|r||}
\hhline{|t:=:t:===:t|}
\multicolumn{1}{||c||}{matrix}
& \multicolumn{3}{c||}{accuracy in bits}\\
\hhline{||~||---||}
 \multicolumn{1}{||c||}{size}  & \rule{0mm}{4mm} 4096 & 8192  & 16384 \\
 \hhline{|:=::===:|}
\rule{0mm}{5mm}50 & 7 & 28 & 111 \\
100 & 29  & 112 & 441 \\
250 & 185 & 441 & 1728 \\
500 & 744 & 3017  & 12262 \\
1000 & 3175 & 12732 & 50674 \\
\hhline{|b:=:b:===:b|}
\end{tabular}
\end{center}
\end{table}
\renewcommand{\baselinestretch}{2}

\renewcommand{\baselinestretch}{1}
\begin{table}[!hp]
\begin{center}
\caption{The mean execution time (sec) of Gauss elimination algorithm on an  Intel  E5-2670 CPU using arbitrary precision library \texttt{GMPLIB}, version 5.0.5 \cite{GMPLIB}. One can observe that the CPU time is increased by about 2.8 when doubling the accuracy, consistent with the complexity of Karatsuba multiplication algorithm. } \label{tab2}
\begin{tabular}{||r||r|r|r||}
\hhline{|t:=:t:===:t|}
\multicolumn{1}{||c||}{matrix}
& \multicolumn{3}{c||}{accuracy in bits}\\
\hhline{||~||---||}
 \multicolumn{1}{||c||}{size}   & \rule{0mm}{4mm}  4096 & 8192 & 16384 \\
 \hhline{|:=::===:|}
\rule{0mm}{5mm}100 & 2.4  & 6.7 & 18 \\
250 & 39 & 107 & 292 \\
500 & 306 & 850  & 2365 \\
1000 & 2384 & 6627 & 18356 \\
\hhline{|b:=:b:===:b|}
\end{tabular}
\end{center}
\end{table}
\renewcommand{\baselinestretch}{2}

\renewcommand{\baselinestretch}{1}
\begin{table}[!p]
\begin{center}
\caption{The  execution time (sec) of Gauss elimination algorithm on a cluster of   Intel  E5-2670 CPUs using arbitrary precision library \texttt{GMPLIB}, version 5.0.5 with accuracy 32768 bits  and MPI interface. We observe CPU time growth with $N$ as $O(N^3)$, and nearly linear reduction of wall time with the number of processes $T$. } \label{tab3}
\begin{tabular}{||r||r|r|r||}
\hhline{|t:=:t:===:t|}
\multicolumn{1}{||c||}{matrix}
& \multicolumn{3}{c||}{number of processes}\\
\hhline{||~||---||}
 \multicolumn{1}{||c||}{size}  & \rule{0mm}{4mm}  36 & 72 & 144 \\
 \hhline{|:=::===:|}
\rule{0mm}{5mm}2000 & 18063 & 7610 & 3565 \\
4000 & 114012 & 57114 & 28521 \\
12000 & -- & -- & 654120 \\
\hhline{|b:=:b:===:b|}
\end{tabular}
\end{center}
\end{table}
\renewcommand{\baselinestretch}{2}

%\end{document}

\end{document}